\documentclass[11pt,a4paper]{article}
\pdfoutput=1

\usepackage{jheppub}

\usepackage{amsmath,amssymb}
\usepackage{graphicx}

\arxivnumber{1111.3952}

\title{Non-Abelian family symmetries in Pati-Salam unification}
\author[a]{Ivo de Medeiros Varzielas}
\emailAdd{ivo.de@udo.edu}
\affiliation[a]{Fakult\"{a}t f\"{u}r Physik, Technische Universit\"{a}t Dortmund
D-44221 Dortmund, Germany}


\abstract{
We present a framework of underlying $SU(3) \times SU(3)$ family symmetries consistent with Pati-Salam unification and discuss advantages that can justify introducing multiple non-Abelian factors. Advantages include improved vacuum alignment and increased predictivity. We explore in this framework deviations from tri-bi-maximal neutrinos, such as relatively large $\theta_{13}$.}

\begin{document}

\maketitle

\section{Introduction \label{Intro}}

Non-Abelian Family Symmetries (FSs) are useful for addressing the flavour problem of the Standard Model (SM) and beyond - they can order the SM fermion masses and mixing and also alleviate flavour issues of SM extensions such as the SUSY flavour problem \cite{Oscar1, Oscar2, Graham} or the flavour problem of Multi Higgs Doublet Models \cite{IvoMHDM}.

Usually non-Abelian FSs are used in conjunction with one or more Abelian symmetries. We consider in some detail how introducing more than one non-Abelian symmetry can lead to several advantages in order to motivate their use. We present specific models in order to make readily apparent that just as a single non-Abelian symmetry can produce relations between different generations, using more than one can enable further control over the mass structures. In \cite{IvoMHDM} a toy model exemplified quite clearly how two $SU(3)$ factors would forbid undesirable Yukawa terms that with a single non-Abelian symmetry could not be disallowed (regardless of any extra Abelian auxiliary symmetries employed). Examples of works using two non-Abelian family symmetries for other purposes include \cite{Appelquist:2006ag, Sumino:2008hy, Koide:2010hp}.

The relative complexity of simultaneously addressing all the fermions structures by implementing FSs together with Grand Unified Theories (GUTs) makes any advantages that come from using more non-Abelian symmetries particularly worth investigating. It also may be that such models can more easily accommodate the values of $\theta_{13}$ pointed at by results from T2K \cite{Abe:2011sj}, see also e.g. the recent global fits \cite{Fogli:2011qn, Schwetz:2011zk}. These models are also expected to have higher predictivity through reduced number of accidental terms and one can make use of the symmetries to address technical issues regarding the vacuum alignment. All these aspects are studied in the following sections.


As we are interested in scenarios with FS SUSY GUTs, we consider the $SU(4) \times SU(2)_L \times SU(2)_R$ Pati-Salam (PS) framework. Arguably complete $SO(10)$ unification is more appealing, but the maximal FS commuting with the $SO(10)$ GUT is a single $SU(3)$ and furthermore requiring straightforward unification into $SO(10)$ is highly constraining - see e.g. \cite{Ivo1, Ivo3} for examples. PS is extremely appropriate for our current purpose as the maximal FS commuting with the gauge group is $SU(3)\times SU(3)$. A Minimal Flavour Violation approach to a PS GUT would therefore consider that FS at the spurion level, which is another motivation to consider specific realisations at the familon level. Due to the Left-Right (LR) structure of PS, one can even imagine that the $SU(2)_L$ and $SU(2)_R$ factors of the GUT could originate together with the $SU(3)_{LF}\times SU(3)_{RF}$ of the FS from a sufficiently higher rank GUT which breaks first into $SU(5)_{L} \times SU(5)_{R}$ factors - although that is quite beyond the scope of this work and so we do not concern ourselves with the possible origins of the FS further.

\section{Pati-Salam with $SU(3) \times SU(3)$}

\subsection{Vacuum alignment \label{sec:VEVs}}

The results in sections \ref{sec:exact} and \ref{sec:deviations} entirely depend on a particular vacuum expectation value (VEV) structure to be obtained for the familons (PS gauge group singlets). Their VEVs must align along specific directions and we require mild hierarchies between some of their magnitudes. The different VEV magnitude hierarchies can be obtained by radiative corrections driving the masses of the familons negative at slightly different scales.

We now argue that the specific directions required can be obtained from the family symmetries employed: although the framework is based on underlying $SU(3)_{LF}\times SU(3)_{RF}$ FSs, the vacuum alignment will proceed with discrete non-Abelian subgroups and arises quite naturally through soft SUSY breaking terms not allowed by $SU(3)$ as originally suggested in \cite{Ivo3}. We start by discussing the required VEVs and breaking pattern still referring to $SU(3)_{LF}\times SU(3)_{RF}$. These desired directions are particularly relevant for the mixing of the lepton sector, but as we are in a unified framework all fermions structures are derived from them. The first stage of breaking occurs simultaneously, as in order to more naturally have the third generation strongly hierarchical we consider one LR familon charged under both groups: $\phi_{33}$. This enables a fermion mass term at lower order (this desirable feature was addressed recently in the context of GUT models by the use of larger representations of a single non-Abelian FS \cite{KingPSL1, KingPSL2}, c.f. \cite{Ivo3}). When the mass term of $\phi_{33}$ becomes negative it breaks the FSs by acquiring a VEV. If the FSs were continuous then it would merely be a basis choice to designate this direction as the $(0,0,1)$ direction under both groups, but as the alignment comes about through a discrete group this direction arises from terms that depend not just on the magnitude of the VEV but also on its particular direction. As discussed further, it is possible to obtain $(0,0,1)$ up to permutation of the non-zero value, therefore that it would be the third under both groups is merely a re-labeling of the third direction without loss of generality.
At this initial stage of breaking, the pseudo-familon (not strictly a familon as it is a PS non-singlet) $\theta$ that is charged under the $SU(3)_{RF}$ must also acquire its VEV in the same direction. The differences between the L and R sectors appear at the next stages of breaking: the dominant L familon $\phi_{123}$ acquires the largest L VEV in the $(1,1,1)$ direction of $SU(3)_{LF}$, whereas the next stage of $SU(3)_{RF}$ breaking is attributed to $\phi_{2}$ that goes in the $(0,1,0)$ direction (orthogonal to the existing $\theta$ VEV). The last L familon $\phi_{23}$ acquires a $(0,1,-1)$ VEV i.e. orthogonal to the previous direction and analogously this happens for the last R familon $\phi_{1}$ that develops along $(1,0,0)$.

In order to actually align along these directions, we rely on a suitable discrete subgroup of each $SU(3)$ in analogy with \cite{Ivo3} (see also \cite{Dphil}). $\Delta(27)$ (used in the original suggestion) is a small subgroup of $SU(3)$ with distinct triplet and anti-triplet representations (although as pointed out in \cite{Luhn:2007sy}, the semi-direct product of cyclic groups $C_7 \rtimes C_3$ is a smaller subgroup of $SU(3)$ and still includes the necessary ingredients). Regardless of the specific subgroup chosen for each $SU(3)$, the discrete alignment mechanism works without upsetting the $SU(3)$ invariance of the Yukawa and Majorana terms: the messengers involved in each sector have different quantum numbers and therefore it is possible to enable the relevant $SU(3)$-breaking invariants in the soft SUSY breaking potential terms that align the VEVs. The $SU(3)$ non-invariant terms appear in the potential arising from components of D-terms such as $\chi \chi ( \phi \phi^\dagger  \phi \phi^\dagger)$ (where $\chi$ communicates SUSY breaking). The messengers involved that allow this type of term may have masses close to the Planck mass so the resulting terms are expected to be rather small e.g. suppressed by the gravitino mass over the Planck mass. But even though these $SU(3)$-breaking discrete invariants are tiny they are non-vanishing, their presence distinguishes directions and is enough to align VEVs (note that the $SU(3)$ continuous invariants are unable to discriminate absolute direction of VEVs, they can at most do relative alignments). For a single familon $\phi$, the potential would include quartic $\phi^i \phi^\dagger_i \phi^i \phi^\dagger_i$ (there is an implicit sum over the repeated generation index $i$). Depending on the sign of the coefficient in front this term naturally results either in the $(0,0,1)$ direction (negative sign) or in the $(1,1,1)$ direction (positive sign). As we employ multiple familons, we discuss the interplay between the possible quartics of this type.

At this level the double non-Abelian framework has technical advantages. In the original mechanism \cite{Ivo3} there was some tension with obtaining the $(0,0,1)$ together with the $(1,1,1)$ direction - they are natural for a single field and by extension most natural for a single dominant VEV. We want hierarchical third generations, so with implicit simultaneous sum over $i$ and $j$:
\begin{equation}
V_{33} \propto - \phi^{ij}_{33} \phi^\dagger_{33_{ij}} \phi^{ij}_{33} \phi^\dagger_{33_{ij}}/M^4 
\end{equation}
In this expression $i$ is an $SU(3)_{LF}$ generation index and $j$ an $SU(3)_{RF}$ generation index, with $M$ generically denoting the messenger mass involved in the term.
By having two discrete subgroups it is natural to have the dominant LR $\phi_{33}$ VEV separate and unable to interfere with the alignment term of $\phi_{123}$ so in the alignment sense we have another dominant VEV:
\begin{equation}
V_{123} \propto + \phi^i_{123} \phi^\dagger_{123_i}  \phi^i_{123} \phi^\dagger_{123_i}/M^4
\end{equation}
That the double groups separate the familon alignment naturally can be seen explicitly by considering distinct types of messengers for each type of familon. Familons carrying just one index (e.g. $\phi^i_{123}$) contract with messengers charged only under that family subgroup (e.g. $A^i$, $\bar{A}_i$) in tri-linear discrete invariants with a sum over the repeated index (e.g. $A^i \phi^i_{123} A^i$), whereas $\phi_{33}$ has a similar tri-linear discrete invariant that repeats both indices with messengers charged under both FSs (e.g. $\alpha^{ij} \phi_{33}^{ij} \alpha^{ij}$). Each messenger pair has its own mass term (e.g. $A^i \bar{A}_i$) which gives meaning to the generic $M$ suppressions shown so far. With this messenger structure the repeated indices are carried through the four insertions of the familon. In order for each type of familon to be secluded from the other types it is important that e.g. $\phi^\dagger_{33_ij} \phi^{ij}_{33} \phi^\dagger_{123_i} \phi^i_{123}$ is not generated, so messengers such as $\bar{B}_j$ in the tri-linear coupling $\phi^{ij}_{33} \bar{A}_i \bar{B}_j$ must be very massive or absent from the complete theory in order to preserve the alignment.
For the same type of reason, $\phi_{123}$ does not interfere with $\phi_{2}$ alignment:
\begin{equation}
V_{2} \propto - \phi^\dagger_{2_i} \phi^i_{2} \phi^\dagger_{2_i} \phi^i_{2}/M^4
\end{equation}
$\phi_{23}$ and $\phi_{1}$ can be aligned by one of the possible mixed terms that with a positive sign imposes orthogonality with respect to the dominant VEVs of their type (note the $SU(3)$ invariant contractions of the indices):
\begin{equation}
V_{23} \propto + \phi^i_{23} \phi^\dagger_{123_i} \phi^j_{123} \phi^\dagger_{23_j}/M^4
\end{equation}
\begin{equation}
V_{1} \propto + \phi^i_{1} \phi^\dagger_{2_i} \phi^j_{2} \phi^\dagger_{1_j}/M^4 
\end{equation}
Due to the expected hierarchy in VEV magnitudes these should naturally be sub-leading to the previous terms but leading over quartics involving just $\phi_{23}$, which would prefer to align its VEV along a different direction.
One issue still needs to be addressed as $V_{23}$ leads to a complex orthogonality between the fields, when we require that these two VEVs produce orthogonal eigenvectors. The terms discussed so far do not favour specific phase directions but we can address this issue with the term:
\begin{equation}
V_{p} \propto - (\phi^i_{123} \phi^i_{123} \phi^i_{123}) (\phi^\dagger_{123_j} \phi^\dagger_{123_j}\phi^\dagger_{123_j})/M^6
\end{equation}
This is a higher order term with two additional insertions that involves the $\Delta(27)$ invariant summing over three (anti-)triplet indices (without the Levi-Civita tensor). $V_p$ discriminates between VEV phases. Some of the minima within the set of degenerate minima that this term creates have no relative phase between the second and third component. In this sense this term can make the $\phi_{123}$ VEV effectively real and together with $V_{23}$ we obtain the desired kind of orthogonality with $\phi_{23}$.

A noteworthy advantage of having more than one non-Abelian factor is that it naturally enables the $(1,1,1)$ VEV to be the most dominant of its type for alignment purposes while not needing to be the most dominant in the mass structures (in \cite{Ivo3} an additional dominant $(1,1,1)$ VEV was required as well as the  smaller $(1,1,1)$ VEV that participated in the mass structures). Considering in particular how the SUSY flavour problem can be addressed by FSs there is the concern that larger $(1,1,1)$ VEVs can cause issues through K\"ahler corrections \cite{Oscar1, Oscar2}, and so it is quite relevant that this issue can be mitigated in this framework with multiple non-Abelian factors.

\subsection{Exact tri-bi-maximal neutrinos \label{sec:exact}}

In relation to \cite{Ivo1, Ivo3} we have abandoned the extremely constraining requirement of straightforward $SO(10)$ unification which would require the L and R sectors to transform in the same way under the FS. The models presented here are more directly comparable to \cite{KingSO3} in terms of VEVs used and respective mass structures.
We start by considering models which aim to get exact tri-bi-maximal (TB) neutrino eigenstates. Leptonic mixing then gets perturbed from TB due to charged lepton corrections that are related to the quark sector due to the GUT relations (as in the GUT FS models discussed above). We can therefore already state that the leptonic mixing angles predicted in this class of models are rather close to the TB values, and namely that with $\theta_{C}$ the Cabibbo angle (from the quark sector): 
\begin{equation}
\theta_{13} \simeq \theta_{C}/(3 \sqrt{2}) \simeq 3^o
\label{eq:t13GUT}
\end{equation}

For the purpose of achieving exact neutrino TB with fermion structures similar to those in \cite{KingSO3}, in our current framework it is sufficient to use a single Abelian factor $U(1)_F$ (c.f. \cite{KingSO3}). As all familons discussed so far are anti-triplets, it is not possible to form combinations carrying only non-trivial $U(1)_F$ charge and for that reason we introduce one extra familon $\sigma$. In terms of notation, $\Psi_i$ are the L fermions and $\Psi^c_j$ the conjugates of the R fermions. $H$ are the SM or MSSM-like Higgs fields (as this is a SUSY construction we require two doublets) and $\Sigma$ is a Georgi-Jarlskog field - a scalar in a larger representation of the GUT that develops a VEV in a R hypercharge direction, notably one that vanishes in the R neutrino direction (see e.g. \cite{Dphil} for more details). Unlike in section \ref{sec:VEVs}, for the discussion of terms that give rise to the mass structures in sections \ref{sec:exact} and \ref{sec:deviations} we consistently use $i$ as an $SU(3)_{LF}$ generation index and $j$ and $k$ as $SU(3)_{RF}$ generation indices.

\begin{table}
\begin{center}
\begin{tabular}{|c|ccc|cc|c|}
\hline
Field & $SU(4)$ & $SU(2)_{L}$ & $SU(2)_{R}$ & $SU(3)_{LF}$ & $SU(3)_{RF}$ & $U(1)_{F}$\\ \hline
$\psi$ & $4$ & $2$ & $1$ & $3$ & $1$ & $0$\\
$\psi^c$ & $\bar{4}$ & $1$ & $2$ & $1$ & $3$ & $0$\\ 
$\theta$ & $4$ & $1$ & $2$ & $1$ & $\bar{3}$ & $0$\\ \hline
$H$ & $1$ & $2$ & $2$ & $1$ & $1$ & $0$\\
$\Sigma$ & $15$ & $1$ & $3$ & $1$ & $1$ & $-1$ \\ \hline
$\phi_{33}$ & $1$ & $1$ & $1$ & $\bar{3}$ & $\bar{3}$ & $0$\\ 
$\phi_{123}$ & $1$ & $1$ & $1$ & $\bar{3}$ & $1$ & $-2$\\ 
$\phi_{23}$ & $1$ & $1$ & $1$ & $\bar{3}$ & $1$ & $1$\\ 
$\phi_{2}$ & $1$ & $1$ & $1$ & $1$ & $\bar{3}$ & $0$\\
$\phi_{1}$ & $1$ & $1$ & $1$ & $1$ & $\bar{3}$ & $-1$\\
$\sigma$ & $1$ & $1$ & $1$ & $1$ & $1$ & $2$ \\ \hline
\end{tabular}
\caption{Field and symmetry content of a model with exact TB neutrinos. \label{ta:U1}}
\end{center}
\end{table}

With the symmetries of Table \ref{ta:U1} we obtain the following Yukawa superpotential:
\begin{equation}
P_{Y} \sim \Psi_i \Psi_j^c H \left[ \frac{\phi_{33}^{ij}}{M} + \frac{\phi_{23}^{i} \phi_{1}^{j}}{M^2} + \frac{\phi_{23}^{i} \phi_{2}^{j} \Sigma}{M^3} + \frac{\phi_{123}^{i} \phi_{2}^{j} \sigma}{M^3} + \frac{\phi_{123}^{i} \phi_{1}^{j} \Sigma \sigma \sigma}{M^5} \right]
\end{equation}
The RH Majorana masses arise through $\theta$:
\begin{equation}
P_{M} \sim \Psi^c_j \Psi^c_k \frac{\theta^{j}\theta^{k}}{M} + \Psi^c_j \Psi^c_k \theta^{j} \phi_1^k \sigma \frac{ \left(\theta \phi_2 \phi_1 \right)}{M^5}
 \end{equation}
\begin{equation}
+\Psi^c_j \Psi^c_k \sigma \frac{ \left(\theta \phi_2 \phi_1 \right)}{M^3} \frac{ \left(\theta \phi_2 \phi_1 \right)}{M^3} \left[\frac{\phi_{1}^{j} \phi_{1}^{k} \sigma}{M^3} + \frac{\phi_{2}^{j} \phi_{2}^{k}}{M^2} \right]
\end{equation}
$\left(\theta \phi_2 \phi_1 \right)$ is the $SU(3)_{RF}$ invariant involving the antisymmetric Levi-Civita contraction. The relevant structure for TB is inside the square brackets of the last line and $M$ generically denotes the messenger masses which are not the same in each sector (and in particular unrelated with the ones mentioned in the discussion of the alignment).

These fermion superpotentials are phenomenologically viable provided the VEVs have a moderate hierarchy:
\begin{equation}
\langle \Sigma \rangle/M \sim \epsilon^{1/2} > \langle \sigma \rangle/M \sim \epsilon
\end{equation}
\begin{equation}
\label{eq:LVEVs}
\langle \phi_{123} \rangle/M \sim \epsilon^{3/2} > \langle \phi_{23} \rangle/M \sim \epsilon^2
\end{equation}
\begin{equation}
\langle \phi_{2} \rangle/M \sim \epsilon^{1/2} > \langle \phi_{1} \rangle/M \sim \epsilon
\end{equation}
In these symbolic relations we refer to the magnitudes of the non-zero entries of the VEVs. $\epsilon \simeq 0.15$ is the expansion parameter when the messenger masses considered are the ones that generate the down and charged lepton masses - when taking ratios of two VEVs the messenger mass would cancel if included in both and so $\epsilon$ is useful for comparing the VEV magnitudes (see \cite{KingSO3, Ivo1, Dphil} for more details).
These hierarchies between the VEVs are not very strict as there are several $O(1)$ parameters involved - what is important is that charged fermion textures similar to those in \cite{KingSO3} are produced with the term $\phi_{23} \phi_2 \Sigma/M^3$ appearing at about $\epsilon^2$, the $\phi_{23} \phi_1/M^2$ and $\phi_{123} \phi_2 \sigma/M^3$ at about $\epsilon^3$. Denoting as accidental terms any allowed terms that are not necessary to achieve the desired fermion textures, there is (only) one accidental Yukawa term $\phi_{123} \phi_1 \Sigma \sigma \sigma/M^5$ and it is sufficiently suppressed (in this case at about $\epsilon^5$). Eq.(\ref{eq:LVEVs}) is consistent with the discussion of section \ref{sec:VEVs} as $\phi_{123}$ is the dominant of the L familons (note the alignment terms enter with the square of the fields involved so the hierarchies are exacerbated).

With respect to the neutrinos, Sequential Dominance (SD) takes place through type I seesaw as in \cite{KingSO3} to produce TB neutrino mixing (the already discussed deviations from TB on leptonic mixing then result from the Cabibbo angle and lead to eq.(\ref{eq:t13GUT})). Due to SD the dominant Yukawa term with $\phi_{33}$ gets effectively erased after seesaw due to the huge hierarchy of $\theta \Psi^c \theta \Psi^c $ relative to the other Majorana terms. This also means the light neutrinos necessarily have a normal hierarchy with the lightest mass state being approximately massless. We can then check that the atmospheric and solar eigenstates constrain the model:
\begin{equation}
\sqrt{\frac{\Delta m_{\odot}^2}{\Delta m_{@}^2}} \simeq \epsilon \sim \frac{\langle \sigma \rangle}{M_N} \left( \frac{\langle \sigma \rangle}{M_\nu} \right)^2 \frac{\langle \phi_{123} \rangle^2}{\langle \phi_{23} \rangle^2}
\label{eq:massratio}
\end{equation}
This relation is only valid up to the $O(1)$ parameters of the two Yukawa and two Majorana terms involved but it is possible to see this as a consistency test of the model: the hierarchy between the squared mass differences must be reproduced and in this construction the relative magnitude of $\langle \phi_{123} \rangle$ with $\langle \phi_{23} \rangle$ and the VEV of $\sigma$ can combine to do so consistently with eq.(\ref{eq:LVEVs}). In the case that the dominant neutrino messengers correspond to the R sector, their generic Yukawa messenger mass is the same as the generic Majorana messenger mass ($M_\nu \sim M_N$) and this relation would constrain the magnitude of $\langle \sigma \rangle/M_{N}$.

One can change the $U(1)_F$ factor to a sufficiently large $C_N$ group such that no dangerous accidental terms are allowed. $C_6$ is already safe producing only the exact same terms allowed by $U(1)_{F}$ (and inevitably all those terms repeated with multiples of $\sigma^3$) - i.e. the leading terms are the same. The $C_5$ case is interesting and will be discussed in the following section.

The power of the underlying $SU(3)\times SU(3)$ framework manifests itself on the Yukawa sector by allowing only L (R) familons to contract with $\Psi$ ($\Psi^c$) and is apparent when comparing to the symmetries and accidental terms of \cite{KingSO3, Ivo1, Ivo3} - although of course one should note that \cite{Ivo1, Ivo3} straightforwardly embed into $SO(10)$ GUTs and that even though they have more complicated Abelian factors they have only a single non-Abelian factor.

\subsection{Deviations from tri-bi-maximal neutrinos \label{sec:deviations}}

Recent observations point to $\theta_{13}$ being relatively large. This creates tension with the vanishing value predicted by exact TB and by extension also with the non-zero but relatively small angle predicted in the class of FS GUT models which include the $U(1)_F$ (or $C_6$) model discussed in section \ref{sec:exact}. Nonetheless it should be clear that non-Abelian symmetries remain extremely appealing and in particular discrete symmetries may naturally produce interesting mixing patterns in accordance with the $\theta_{13}$ values indicated in \cite{Abe:2011sj, Fogli:2011qn, Schwetz:2011zk}, see e.g. \cite{Claudiat13} \footnote{Recently \cite{Rashed:2011xe} also tackled deviations from TB without using non-Abelian family symmetries.}. In \cite{Kingt13} TB was abandoned but tri-maximal mixing was kept. We consider that TB can remain as an excellent approximation to the neutrino data and interpreted as a strong hint of an underlying FS arranging for TB at some level - particularly in such GUT frameworks where the charged lepton corrections already provide a source of TB deviations, one can imagine an appealing situation where one starts with approximate TB neutrinos and can then add the small Cabibbo angle corrections that arise through the charged leptons due to the GUT relations between the fermions. The framework presented here (with the power of the non-Abelian groups) is ideal to demonstrate this.
There are essentially three distinct sources for introducing a perturbation to the neutrino mixing in this class of models - in the VEV structure itself, in the Majorana structure, or in the Yukawa structure (of course one could have more than one of these effects operating). 
Altering the VEV structure would be done in a way similar to \cite{King:2009qt}.
Deviations in the Majorana structure are expected to be invisible to experimental tests so in theory a model featuring them as the source of deviation of exact neutrino TB could always be invoked to provide a larger value of $\theta_{13}$. In contrast perturbations in the Yukawa structure also appear in the charged fermion textures due to the GUT relations, and therefore this possibility is more appealing as it can in theory maintain some GUT link between the charged fermions and the deviations from TB that produce the larger $\theta_{13}$ angle. We attempt then to obtain deviations arising in the Yukawa structure only.

One such example arises from reconsidering the $U(1)_F$ or $C_6$ model of the previous section, but replacing now the $U(1)_F$ by $C_5$ (instead of by $C_6$). The field content, PS and $SU(3)_{LF}\times SU(3)_{RF}$ assignments of section \ref{sec:exact} are unchanged. For explicitness, the $C_5$ charges are listed in Table \ref{ta:C5}.

\begin{table}
\begin{center}
\begin{tabular}{|c|cccccc|}
\hline
Field & $\phi_{123}$ & $\phi_{23}$ & $\phi_{2}$ & $\phi_1$ & $\Sigma$ & $\sigma$ \\ \hline
$C_5$ & $3$ & $1$ & $0$ & $4$ & $4$ & $2$ \\ \hline
\end{tabular}
\end{center}
\caption{$C_5$ transformations of the fields. \label{ta:C5}}
\end{table}
By construction the Yukawa and Majorana terms listed in the previous section are preserved, but one must add to the Yukawa the following new accidental terms:
\begin{equation}
P_{5} \sim \Psi_i \Psi_j^c H \left[\frac{\phi_{23}^{i} \phi_{2}^{j}\sigma \sigma}{M^4} + \frac{\phi_{23}^{i} \phi_{1}^{j} \Sigma \sigma \sigma \sigma}{M^6} + \frac{\phi_{123}^{i} \phi_{1}^{j} \sigma \sigma \sigma \sigma}{M^6} + \frac{\phi_{123}^{i} \phi_{2}^{j} \Sigma \sigma \sigma \sigma \sigma}{M^7} \right]
\end{equation}
The only significant term is the first one, which enters at about $\epsilon^4$ in the second column of the Yukawa matrix and perturbs what previously became the neutrino eigenstate proportional to the $(1,1,1)$ direction - which itself enters at about $\epsilon^3$. This is a clear example of a correction to exact TB neutrinos arising from only one significant source in the neutrino Yukawa sector.

As we would like to generate a $\theta_{13}$ angle around the central values from \cite{Abe:2011sj, Fogli:2011qn, Schwetz:2011zk} we consider that an $(\epsilon, 1, -1)$ eigenstate produces just that order of magnitude for the angle - with the same $\epsilon$ numerical value as used for some of the fermion hierarchies and ratios of VEVs \footnote{We acknowledge Graham Ross for this interesting observation.}.
We searched for concrete examples where we generate such an entry through $\Psi_i \Psi_j^c H \phi_{123}^{i} \phi_{1}^{j}/M^2$ with an appropriate suppression due to $\sigma/M$ - ideally with the following three conditions, similarly to the previous $C_5$ example: 1. no other terms produce significant deviations from TB; 2. with appropriate VEV hierarchies such that the fit works with natural $O(1)$ coefficients; 3. without adding extra fields. We were unable to verify all those conditions due to correlations between the Abelian charges of the familons. As some examples that relax one of conditions, we list the Abelian charges of models 1Y, 1M, 2 and 3 in Table \ref{ta:MF}. The field content, PS and $SU(3)_{LF}\times SU(3)_{RF}$ assignments of section \ref{sec:exact} are the same except for model 3 which has added $\sigma'$ (which is only charged non-trivially under $U(1)_3$, like $\sigma$). 

\begin{table}
\begin{center}
\begin{tabular}{|c|ccccccc|}
\hline
Field & $\phi_{123}$ & $\phi_{23}$ & $\phi_{2}$ & $\phi_1$ & $\Sigma$ & $\sigma$ & $\sigma'$ \\ \hline
1Y: $C_N$ & $N/4$ & $N/2$ & $0$ & $N/2$ & $N/2$ & $3N/4$ & $\ast$ \\ 
1M: $U(1)_{1M}$ & $-1$ & $1$ & $0$ & $-1$ & $-1$ & $1$ &  $\ast$ \\ 
2: $C_{N'}$ & $0$ & $N'/2$ & $0$ & $N'/2$ & $N'/2$ & $N'/2$ &  $\ast$ \\
3: $U(1)_{3}$ & $-2$ & $1$ & $0$ & $-1$ & $-1$ & $2$ & $3$ \\ \hline
\end{tabular}
\end{center}
\caption{Abelian charges of models 1Y, 1M, 2 and 3. \label{ta:MF}}
\end{table}

We have found two types of models violating condition 1. Model 1Y is similar to the $C_5$ case discussed already as it adds $\Psi_i \Psi_j^c H \phi_{23}^{i} \phi_{2}^{j} \sigma \sigma/M^4$, but the term $\Psi_i \Psi_j^c H \phi_{123}^{i} \phi_{1}^{j} \sigma \sigma \sigma/M^5$ appears (one order lower than in the $C_5$ model). The would-be texture zero in the neutrino Yukawa structure could be populated, but obviously not as the only significant TB deviation.
Model 1M in contrast has no possible $\Psi_i \Psi_j^c H \phi_{23}^{i} \phi_{2}^{j}/M^2$ term and can populate the texture zero but has the structure $\Psi^c_j \Psi^c_k  \left[\phi_{1}^{j} \phi_{1}^{k} \sigma \sigma/M^4 + \phi_{1}^{j} \phi_{2}^{k} \sigma/M^3 + \phi_{2}^{j} \phi_{2}^{k}/M^2 \right]$ in the Majorana sector (note that only $SU(3)_{RF}$ is at work so we do not benefit from the $SU(3) \times SU(3)$ framework). The off-diagonal $\phi_1 \phi_2$ terms are not suppressed relative to the $\phi_1 \phi_1$ term and will introduce significant effects that spoil TB mixing.

We can simply take model 1Y and 1M and trade their extra terms that significantly deviate from TB for unnaturally suppressing those extra terms with coefficients that are arbitrarily small (instead of naturally $O(1)$), thus violating condition 2. Another example that is interesting due to its significantly different fermion structures is model 2: the Majorana sector $\Psi^c_j \Psi^c_k  \left[\phi_{1}^{j} \phi_{1}^{k}/M^2 + \phi_{1}^{j} \phi_{2}^{k} \sigma/M^3 + \phi_{2}^{j} \phi_{2}^{k}/M^2 \right]$ has $\phi_1 \phi_2$ terms suppressed by $\sigma/M$ that can be made negligible compared to the diagonal terms through the messenger mass of the sector, unlike in model 1M. There are two noteworthy Yukawa terms $\Psi_i \Psi_j^c H \phi_{123}^{i} \phi_{1}^{j} \sigma/M^3$, $\Psi_i \Psi_j^c H \phi_{23}^{i} \phi_{2}^{j} \sigma/M^3$ - at this stage there would already be some fine-tuning if we want to keep the first one and neglect the second one, although not as extreme in the tuned version of model 1Y. Model 2 becomes less appealing when we consider its equivalent to eq.(\ref{eq:massratio}) - interestingly as every relevant term enters at the same order it directly constrains the hierarchy of the L familon VEVs, regardless of messenger masses:
\begin{equation}
\sqrt{\frac{\Delta m_{\odot}^2}{\Delta m_{@}^2}} \simeq \epsilon \sim \frac{\langle \phi_{123} \rangle^2}{\langle \phi_{23} \rangle^2}
\end{equation}
Clearly unlike the model in section \ref{sec:exact} this time the consistency check is not naturally verified as the magnitudes of the VEVs go in the opposite direction to the one we argued is desirable in section \ref{sec:VEVs}. So at least in this context the model firmly belongs in the class of unnatural fine-tuning and strongly violates condition 2.

Finally, it is relatively easy to build a model that works - if we are willing to violate condition 3. Simply introduce another field $\sigma'$ and adjust the Abelian charge and magnitude of $\langle \sigma' \rangle/M$ - the only concern is that it does not enable any other unwanted terms. Model 3 starts with the structure shown in section \ref{sec:exact} and directly enables $\Psi_i \Psi_j^c H \phi_{123}^{i} \phi_{1}^{j} \sigma'/M^3$ (it also enables $\Psi_i \Psi_j^c H \phi_{123}^{i} \phi_{2}^{j} \Sigma \sigma'/M^4$, but that term won't affect the existing structure). In this particular construction we also get a term in the Majorana sector that constrains the respective messenger mass against the hierarchy of the VEVs, $\langle \sigma ^2 \rangle/M_N^2 > \langle \sigma' \rangle/M_N$. The problem with this approach is that it is not elegant: a field was added just for this purpose - the magnitude of its VEV is a new parameter that directly controls the deviation from TB and ultimately the value of $\theta_{13}$ (note however that we predict the correct order of magnitude as we require the fit to work for the charged fermions, so we can not perturb the previous structures too much).

Before concluding, we note that it may be possible to get these models to work as intended through specific UV completions (model 3 already works due to $\sigma'$). Problematic terms that at the non-renormalisable level could not be suppressed or disallowed by the symmetries may be fixed at the renormalisable level by the field content (the specific messenger structures). Explicit UV completions are well beyond of the scope of the present work, but \cite{IvoLuca} clearly demonstrates this strategy that can greatly increase predictivity, preserving all the necessary terms but with many terms that would otherwise be allowed being absent solely due to the lack of the necessary messengers.

\section{Conclusion}

We have proposed and motivated the use of multiple non-Abelian family symmetries for building models. We explored a Grand Unified Pati-Salam framework with $SU(3)\times SU(3)$ family symmetries. By the use of specific alignment and model examples we illustrated several advantages of this framework: the natural separation of sectors led to an improved alignment mechanism, mitigation of possible SUSY flavour issues, and to a lower order Yukawa term for the third generation of fermions; the increased control over the allowed terms led to a decrease in the Abelian factors in the total family symmetry and to increased predictivity through the absence of accidental terms in general, and in particular proved ideal to explore possible deviations from exact tri-bi-maximal neutrinos.

\acknowledgments

I want to thank Graham Ross for helpful discussions, particularly about the vacuum alignment mechanism.
This work was supported by DFG grant PA 803/6-1.

\bibliographystyle{JHEP}
\bibliography{refs}

\providecommand{\href}[2]{#2}\begingroup\raggedright\begin{thebibliography}{10}

\bibitem{Oscar1}
G.~G. Ross, L.~Velasco-Sevilla, and O.~Vives, {\it {Spontaneous CP violation
  and nonAbelian family symmetry in SUSY}},  {\em Nucl.Phys.} {\bf B692} (2004)
  50--82, [\href{http://xxx.lanl.gov/abs/hep-ph/0401064}{{\tt
  hep-ph/0401064}}].

\bibitem{Oscar2}
L.~Calibbi, J.~Jones-Perez, A.~Masiero, J.-h. Park, W.~Porod, {\em et.~al.},
  {\it {FCNC and CP Violation Observables in a SU(3)-flavoured MSSM}},  {\em
  Nucl.Phys.} {\bf B831} (2010) 26--71,
  [\href{http://xxx.lanl.gov/abs/0907.4069}{{\tt arXiv:0907.4069}}].

\bibitem{Graham}
Z.~Lalak, S.~Pokorski, and G.~G. Ross, {\it {Beyond MFV in family symmetry
  theories of fermion masses}},  {\em JHEP} {\bf 1008} (2010) 129,
  [\href{http://xxx.lanl.gov/abs/1006.2375}{{\tt arXiv:1006.2375}}].

\bibitem{IvoMHDM}
I.~de~Medeiros~Varzielas, {\it {Family symmetries and alignment in multi-Higgs
  doublet models}},  {\em Phys. Lett.} {\bf B701} (2011) 597--600,
  [\href{http://xxx.lanl.gov/abs/1104.2601}{{\tt arXiv:1104.2601}}].

\bibitem{Appelquist:2006ag}
T.~Appelquist, Y.~Bai, and M.~Piai, {\it {Quark mass ratios and mixing angles
  from SU(3) family gauge symmetry}},  {\em Phys. Lett.} {\bf B637} (2006)
  245--250, [\href{http://xxx.lanl.gov/abs/hep-ph/0603104}{{\tt
  hep-ph/0603104}}].

\bibitem{Sumino:2008hy}
Y.~Sumino, {\it {Family Gauge Symmetry as an Origin of Koide's Mass Formula and
  Charged Lepton Spectrum}},  {\em JHEP} {\bf 05} (2009) 075,
  [\href{http://xxx.lanl.gov/abs/0812.2103}{{\tt arXiv:0812.2103}}].

\bibitem{Koide:2010hp}
Y.~Koide, {\it {Yukawaon Model with $U(3) \times O(3)$ Family Symmetries}},
  {\em J. Phys.} {\bf G38} (2011) 085004,
  [\href{http://xxx.lanl.gov/abs/1011.1064}{{\tt arXiv:1011.1064}}].

\bibitem{Abe:2011sj}
{\bf T2K} Collaboration, K.~Abe {\em et.~al.}, {\it {Indication of Electron
  Neutrino Appearance from an Accelerator-produced Off-axis Muon Neutrino
  Beam}},  {\em Phys. Rev. Lett.} {\bf 107} (2011) 041801,
  [\href{http://xxx.lanl.gov/abs/1106.2822}{{\tt arXiv:1106.2822}}].

\bibitem{Fogli:2011qn}
G.~L. Fogli, E.~Lisi, A.~Marrone, A.~Palazzo, and A.~M. Rotunno, {\it {Evidence
  of theta(13)>0 from global neutrino data analysis}},  {\em Phys. Rev.} {\bf
  D84} (2011) 053007, [\href{http://xxx.lanl.gov/abs/1106.6028}{{\tt
  arXiv:1106.6028}}].

\bibitem{Schwetz:2011zk}
T.~Schwetz, M.~Tortola, and J.~W.~F. Valle, {\it {Where we are on
  $\theta_{13}$: addendum to 'Global neutrino data and recent reactor fluxes:
  status of three- flavour oscillation parameters'}},  {\em New J. Phys.} {\bf
  13} (2011) 109401, [\href{http://xxx.lanl.gov/abs/1108.1376}{{\tt
  arXiv:1108.1376}}].

\bibitem{Ivo1}
I.~de~Medeiros~Varzielas and G.~G. Ross, {\it {SU(3) family symmetry and
  neutrino bi-tri-maximal mixing}},  {\em Nucl.Phys.} {\bf B733} (2006) 31--47,
  [\href{http://xxx.lanl.gov/abs/hep-ph/0507176}{{\tt hep-ph/0507176}}].

\bibitem{Ivo3}
I.~de~Medeiros~Varzielas, S.~F. King, and G.~G. Ross, {\it {Neutrino
  tri-bi-maximal mixing from a non-Abelian discrete family symmetry}},  {\em
  Phys. Lett.} {\bf B648} (2007) 201--206,
  [\href{http://xxx.lanl.gov/abs/hep-ph/0607045}{{\tt hep-ph/0607045}}].

\bibitem{KingPSL1}
S.~F. King and C.~Luhn, {\it {A New family symmetry for SO(10) GUTs}},  {\em
  Nucl.Phys.} {\bf B820} (2009) 269--289,
  [\href{http://xxx.lanl.gov/abs/0905.1686}{{\tt arXiv:0905.1686}}].

\bibitem{KingPSL2}
S.~F. King and C.~Luhn, {\it {A Supersymmetric Grand Unified Theory of Flavour
  with PSL(2)(7) x SO(10)}},  {\em Nucl.Phys.} {\bf B832} (2010) 414--439,
  [\href{http://xxx.lanl.gov/abs/0912.1344}{{\tt arXiv:0912.1344}}].

\bibitem{Dphil}
I.~de~Medeiros~Varzielas, {\it {Family symmetries and the origin of fermion
  masses and mixings}},  \href{http://xxx.lanl.gov/abs/0801.2775}{{\tt
  arXiv:0801.2775}}.

\bibitem{Luhn:2007sy}
C.~Luhn, S.~Nasri, and P.~Ramond, {\it {Tri-Bimaximal Neutrino Mixing and the
  Family Symmetry $Z_7 \rtimes Z_3$}},  {\em Phys. Lett.} {\bf B652} (2007)
  27--33, [\href{http://xxx.lanl.gov/abs/0706.2341}{{\tt arXiv:0706.2341}}].

\bibitem{KingSO3}
S.~King, {\it {Predicting neutrino parameters from SO(3) family symmetry and
  quark-lepton unification}},  {\em JHEP} {\bf 0508} (2005) 105,
  [\href{http://xxx.lanl.gov/abs/hep-ph/0506297}{{\tt hep-ph/0506297}}].

\bibitem{Claudiat13}
R.~d.~A. Toorop, F.~Feruglio, and C.~Hagedorn, {\it {Discrete Flavour
  Symmetries in Light of T2K}},  {\em Phys. Lett.} {\bf B703} (2011) 447--451,
  [\href{http://xxx.lanl.gov/abs/1107.3486}{{\tt arXiv:1107.3486}}].

\bibitem{Rashed:2011xe}
A.~Rashed, {\it {Deviation from Tri-Bimaximal Mixing and Large Reactor Mixing
  Angle}},  \href{http://xxx.lanl.gov/abs/1111.3072}{{\tt arXiv:1111.3072}}.

\bibitem{Kingt13}
S.~Antusch, S.~F. King, C.~Luhn, and M.~Spinrath, {\it {Trimaximal mixing with
  predicted $\theta_13$ from a new type of constrained sequential dominance}},
  \href{http://xxx.lanl.gov/abs/1108.4278}{{\tt arXiv:1108.4278}}.

\bibitem{King:2009qt}
S.~King, {\it {Tri-bimaximal Neutrino Mixing and theta(13)}},  {\em Phys.Lett.}
  {\bf B675} (2009) 347--351, [\href{http://xxx.lanl.gov/abs/0903.3199}{{\tt
  arXiv:0903.3199}}].

\bibitem{IvoLuca}
I.~de~Medeiros~Varzielas and L.~Merlo, {\it {Ultraviolet Completion of Flavour
  Models}},  {\em JHEP} {\bf 02} (2011) 062,
  [\href{http://xxx.lanl.gov/abs/1011.6662}{{\tt arXiv:1011.6662}}].

\end{thebibliography}\endgroup

\end{document}